\documentclass[aps,prb,twocolumn,floatfix,superscriptaddress]{revtex4-1}
\usepackage[english]{babel}
\usepackage[utf8]{inputenc}
\usepackage[T1]{fontenc}
\usepackage{bbold}
\usepackage{epsfig}
\usepackage{amsmath}
\usepackage{hyphenat}
\usepackage{hyperref}
\usepackage{placeins}
\usepackage{bm}
\usepackage[colorinlistoftodos]{todonotes}

\begin{document}

\title{Degeneracies and fluctuations of N\'e\`el skyrmions in confined geometries}

\author{Rick Keesman}
\affiliation{
    Instituut-Lorentz for Theoretical Physics,
    Leiden University,
    Niels Bohrweg 2,
    NL-2333CA Leiden,
    The Netherlands
}

\author{A.O. Leonov}
\affiliation{
    Zernike Institute for Advanced Materials,
    Groningen University,
    Nijenborgh 4,
    9747 AG Groningen,
    The Netherlands
}

\author{P. van Dieten}
\affiliation{
    Institute for Theoretical Physics and Center for Extreme Matter and Emergent Phenomena,
    Utrecht University,
    Leuvenlaan 4,
    3584 CE Utrecht,
    The Netherlands
}

\author{Stefan Buhrandt}
\affiliation{
    Institute for Theoretical Physics and Center for Extreme Matter and Emergent Phenomena,
    Utrecht University,
    Leuvenlaan 4,
    3584 CE Utrecht,
    The Netherlands
}

\author{G. T. Barkema}
\affiliation{
    Instituut-Lorentz for Theoretical Physics,
    Leiden University,
    Niels Bohrweg 2,
    NL-2333CA Leiden,
    The Netherlands
}
\affiliation{
    Institute for Theoretical Physics and Center for Extreme Matter and Emergent Phenomena,
    Utrecht University,
    Leuvenlaan 4,
    3584 CE Utrecht,
    The Netherlands
}

\author{Lars Fritz}
\affiliation{
    Institute for Theoretical Physics and Center for Extreme Matter and Emergent Phenomena,
    Utrecht University,
    Leuvenlaan 4,
    3584 CE Utrecht,
    The Netherlands
}

\author{R.A. Duine}
\affiliation{
	Institute for Theoretical Physics and Center for Extreme Matter and Emergent Phenomena,
	Utrecht University,
	Leuvenlaan 4,
	3584 CE Utrecht,
	The Netherlands
}

\begin{abstract}
The recent discovery of tunable Dzyaloshinskii-Moriya interactions in layered magnetic materials with perpendicular magnetic anisotropy makes them promising candidates for stabilization and manipulation of skyrmions at elevated temperatures. In this article, we use Monte Carlo simulations to investigate the robustness of skyrmions in these materials against thermal fluctuations and finite-size effects. We find that in confined geometries and at finite temperatures skyrmions are present in a large part of the phase diagram. Moreover, we find that the confined geometry favors the skyrmion over the spiral phase when compared to infinitely large systems. Upon tuning the magnetic field through the skyrmion phase, the system undergoes a cascade of transitions in the magnetic structure through states of different number of skyrmions, elongated and half skyrmions, and spiral states. We consider how quantum and thermal fluctuations lift the degeneracies that occur at these transitions, and find that states with more skyrmions are typically favored by fluctuations over states with less skyrmions. Finally, we comment on electrical detection of the various phases through the topological and anomalous Hall effects. 
\end{abstract}

\pacs{pacs}

\maketitle

\section{Introduction \label{sec:intro}}
A skyrmion is a certain type of topological field configuration which was first introduced in particle physics. It corresponds to a classical stationary solution of the equations of motion with which one can associate a topological invariant and describes the emergence of a discrete particle from a continuous field. \cite{skyrme1} More recently, skyrmions have been considered in quantum Hall devices, \cite{sondhi1} Bose-Einstein condensates, \cite{stoof1,leslie1} and liquid crystals. \cite{bogdanov1, ackerman2014, leonov2014} Magnetic skyrmions were predicted \cite{bogdanov2} and recently observed in bulk materials like $\text{MnSi}$ and $\text{Cu}_{2}\text{OSeO}_{3}$ at low temperatures. \cite{muhlbauer2,pappas1,yu2010} In these materials, bulk inversion symmetry is broken which allows for nonzero Dzyaloshinskii-Moriya (DM) interactions \cite{dzyaloshinsky1,moriya1} that lead to helical, conical, and skyrmionic spin textures depending on the applied magnetic field. \cite{bogdanov2} It was proposed that skyrmions are promising candidates for encoding binary data that allow for high-density and low power consumption magnetic memories due to low critical currents for skyrmion motion and their inherent topological stability. \cite{muhlbauer1,sampaio1,romming2013} In addition to fundamental interest concerning the interplay between topology, geometry and fluctuations, it is therefore relevant to understand the behavior of skyrmions at high temperature and in confined thin-film geometries for realizing spintronic devices.

\begin{figure}[ht] \centering
\epsfig{file=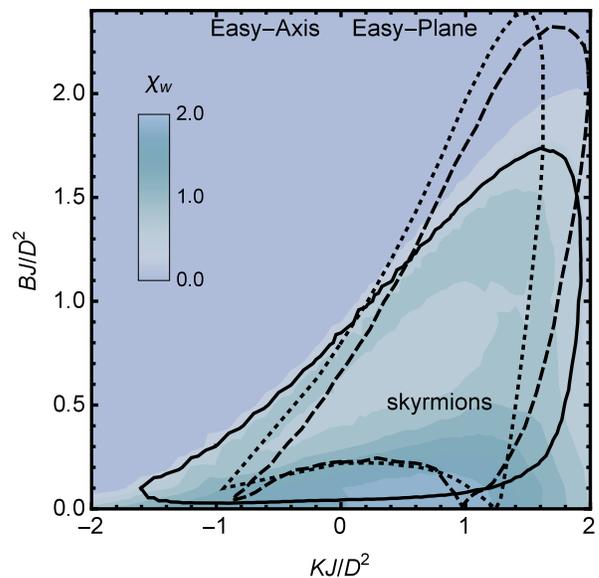,width=0.9\linewidth,clip=}
\caption{The anisotropy-external field phase diagram for a wire geometry  with $L=16$ and $p=8$ at $k_{B} T/J=0.5$ (solid line) and zero temperature (dashed line). The dotted line corresponds to the case of an infinite system at zero temperature.\cite{leonovphd} These lines encircle the region where the winding number is larger than one half. The susceptiblity of the winding number at finite temperature $\chi_w$ is also indicated. }
    \label{fig:phasediagram}
\end{figure}

Motivated by recent experiments on domain wall motion in magnetic thin films with perpendicular magnetic anisotropy (PMA materials) that point to sizeable and tunable DM interactions\cite{emori1,ryu1,ryu2,franken1,je2013,hrabec2014,chen2013} and anisotropy, we use Monte Carlo simulations to investigate the robustness of skyrmions in these systems in confined wire-like geometries against thermal fluctuations. Our main results are:  i) The phase diagram in Fig.~\ref{fig:phasediagram} that shows which spin textures are to be expected at a certain anisotropy strength and applied magnetic field at nonzero temperature. We find that the confining geometry extends (with respect to systems in the thermodynamic limit) the skyrmion phase at the expense of the spiral phase. ii) The cascade of transitions between different magnetic structures that the system undergoes upon lowering the magnetic field through the skyrmion phase, a typical example of which is shown in Fig.~\ref{fig:cascade}.  iii)  The temperature dependence of relative probabilities for occurence of different skyrmions configurations that are degenerate at zero temperature, shown in Fig.~\ref{fig:ratiovst}. We find that at moderate temperature configurations with more skyrmions are typically entropically favored over configuration with less skyrmions. Moreover, we also consider quantum fluctuations and find that these also favor configurations with higher skyrmion number. 

Skyrmions are predicted to occur in several varieties,\cite{bogdanov2} two of which have by now been experimentally identified. One of these is the  N\'e\`el (somtimes also called ``hedgehog'') skyrmion, i.e., a skyrmion in which the magnetization points radially outward from the skyrmion center. The other type of skyrmion (where the magnetization is perpendicular to radii pointing outward from the skyrmion center) is called a Bloch skyrmion.  (See Fig. 1 of Ref.~\onlinecite{sampaio1} for an illustration.) Which of the two types is favored depends on which type of DM interactions are present as dictated by the crystal and/or structural symmetry. In the long-wavelength limit, the DM interactions yield a contribution to the energy density that is a certain combination of so-called Lifshitz invariants, i.e., antisymmetric terms of the form
\begin{equation}
\label{eq:li}
 S_i\frac{\partial S_j}{\partial r} - S_j\frac{\partial S_i}{\partial r}~,
\end{equation} 
with $S_i$ the $i$-th cartesian component of the spin and $r=x,y$ or $z$, a spatial direction.  In MnSi, one of the most-studied skyrmion materials, the DM interactions give a contribution proportional to
\begin{equation}
  {\bf S} \cdot \left( \nabla \times {\bf S}\right),
\end{equation} 
which is straightforwardly shown to be a combination of terms of the form as in Eq.~(\ref{eq:li}), and favors Bloch skyrmions.  For the PMA materials that are the focus of this work, the DM interactions are interface-induced (see also Ref.~\onlinecite{heinze2011}) and stabilize N\'e\`el  skyrmions. They are proportional to the expression
\begin{equation}
\label{eq:dmiinterface}
  ({\bf z} \cdot {\bf S})(\nabla \cdot {\bf S})-({\bf S} \cdot \nabla) ({\bf z} \cdot {\bf S})~,
\end{equation}
and in this particular form are shown to explicitly depend on the symmetry-breaking direction ${\bf z}$ which denotes the normal to the interface.
We note, however, that the above form of DM interactions also arises in crystals of symmetry class $C_{nv}$ and therefore that N\'e\`el skyrmions are stabilized by bulk DM interactions in some materials. This was recently observed in the magnetic semiconductor GaV$_4$S$_8$.\cite{kezsmarki2015} In this material there are no Lifshitz invariants in the $z$-direction and a conical phase is therefore not present. Our results therefore also apply to this case, and, particularly when grown in thin-film form on different substrates, the magnetic anisotropy of this material may be tuned between easy axis and easy plane which enables experimentally exploring the full phase diagram in Fig.~\ref{fig:phasediagram}.

Previous theoretical studies have focused on the skyrmion phase diagram of infinite systems at zero temperature \cite{bogdanov2,randeria1} and nonzero temperature in two and three dimensions. \cite{buhrandt1,han1} The first experimental results on bulk materials \cite{muhlbauer2,pappas1,yu2010} have been extended to confined geometries, such as thin films, e.g. of MnSi \cite{wilson2014} and FeGe. \cite{huang2012} Transitions between states with a different number of skyrmions and other (non-skyrmion) magnetic configurations as a function of field have been discussed experimentally and theoretically (at zero temperature) for thin films of MnSi (and thus for bulk DM interactions leading to Bloch skyrmions) in Refs.~\onlinecite{wilson2012, wilson2013}. Evidence for such a cascade of transitions in magnetoresistance measurements was very recently discussed in Ref.~\onlinecite{du2015}. In these works, the Bloch skyrmion core lines (and field)  lie in the plane of the thin film or along the wire direction due to strong easy-plane anisotropy induced by tensile strain from the substrate. As a result, only in-plane fields stabilize skyrmions whereas fields perpendicular to the thin film (and thus anisotropy plane) lead to conical phases. In the two-dimensional configuration that we consider, however, (see Fig.~\ref{fig:cascade}) the external field is perpendicular to the thin film and skyrmions may be observed for both easy-plane and easy-axis anistropy. For the form of the DM interactions considered here [Eq.~(\ref{eq:dmiinterface})] the conical phase is absent and the easy-plane anisotropy leads to a large external-field range over which skyrmions are stable (see Fig.~\ref{fig:phasediagram}). Moreover, the confined geometry stabilizes skyrmions over spirals as the skyrmion phase becomes larger in the wire geometry with respect to the case of an infinite system. 

Very recently, Du {\it et al.} reported on real-space observation of a cascade of transitions in the magnetic structure of a FeGe wire. \cite{du2015b} Although this system stabilizes Bloch skyrmions, this work gives experimental corroboration of many of our findings, such as the existence of elongated skyrmions, the creation of skyrmions from edge states, and the enhanced stability of skyrmions in the confining geometry. Moreover, these authors also observe that spirals orient themselves with their wave vector along the edge of the wires (see also Ref.~\onlinecite{romming2015}), which is in line with our results as well.  

Finally, we note for completeness that single Bloch skyrmions in nanodisks were considered in Ref.~\onlinecite{beg2013} without taking into account anisotropy and that the effects of thermal fluctuations  and confining geometry on single-skyrmion dynamics was considered in Refs.~\onlinecite{han1,troncoso1, rohart2013}.

The remainder of this article is organized as follows. In Sec.~\ref{sec:model} we discuss our model hamiltonian and the algorithm used in our simulations. Using these simulations we then construct the anisotropy-external field phase diagram for wire-like systems at moderate temperature and discuss the various magnetic phases and cascade of transitions between them in Sec.~\ref{sec:pd}. Hereafter we focus on points in the phase diagram where degeneracies occur at zero temperature and investigate how fluctuations lift these degeneracies. We end with a conclusion and outlook, and briefly discuss electrical detection of the various magnetic configurations. 

\section{Model and simulations \label{sec:model}}
We consider Heisenberg spins ${\bf S}_{\bf r}$ of unit length, where ${\bf{r}}$ denotes the location on a two-dimensional square lattice in the $x$-$y$-plane. The spins interact via a ferromagnetic Heisenberg coupling $J>0$, Dzyaloshinskii-Moriya coupling $D$, and are subject to an anisotropy term $K$ and an external magnetic field ${\bf B}$. The effective hamiltonian is given by
\begin{align}\label{eq:hamiltonian}
H =
	-& J \sum_{\bf r}  {\bf S}_{\bf r}  \cdot \left(  {\bf S}_{{\bf r}+{\bf \hat{x}}} +  {\bf S}_{{\bf r}+{\bf \hat{y}}} \right) \nonumber \\
	+& K \sum_{\bf r}  \left( {\bf S}_{\bf r} \cdot {\bf \hat{z}}  \right)^2 - {\bf B} \cdot \sum_{\bf r}  {\bf S}_{\bf r}  \\
	-& D \sum_{\bf r} \left(  {\bf S}_{\bf r} \times {\bf S}_{{\bf r}+{\bf \hat{x}}} \cdot {\bf \hat{y}} - {\bf S}_{\bf r} \times {\bf S}_{{\bf r}+{\bf \hat{y}}} \cdot {\bf \hat{x}} \right). \nonumber
\end{align}
The above form of the DM interactions (i.e., the term in the hamiltonian proportional to $D$)  corresponds to the discretized version of Eq.~(\ref{eq:dmiinterface}) and is therefore to lowest order in nearest-neighbor coupling appropriate for the PMA materials of interest to us here, i.e., for the situation that the DM interactions arise due to inversion asymmetry induced by the presence of an interface. In the hamiltonian we have neglected dipole-dipole interactions. This is appropriate in the limit of strong DM interactions that lead to small skyrmions and where the dipolar field will only renormalize parameters such as the anisotropy. We note that throughout this article we consider a wire geometry, i.e., a two-dimensional system with periodic boundary conditions in one direction and open boundary conditions in the other direction.  

For simplicity we assume ${\bf B} = B  {\bf \hat{z}}$ perpendicular to the $x$-$y$-plane and take $B>0$ without loss of generality. The dimensionless parameters $k_{B} T / J \equiv 1/(\beta J)$, with $k_B T$ the thermal energy, $B J / D^2$, and $K J / D^2$ determine the state of the system. An important length scale is the pitch length $p$ (in units of the lattice constant) that determines the periodicity of magnetic textures that arise due to competition between DM interaction $D$ and exchange $J$. In case of a spiral state, for example, the pitch length will be the period of the spirals. In the absence of anisotropy, $K=0$, the pitch length is given by \cite{han1} $D/J=\tan \left( 2 \pi / p \right)$  and we will use this definition throughout. This relation can be used to coarse-grain the system by keeping the ratio between the dimensions of the system and the pitch length constant while changing discretization.

We use classical Monte Carlo simulations to sample phase space at finite temperatures $T$. A typical simulation starts with a completely randomized square lattice of spins with  $L \times L$ sites (with one open and one periodic boundary) at fixed parameters $B J / D^2$ and $K J / D^2$ at a scaled temperature $k_{B} T/J=10.0$ far above the critical temperature of the system. After several lattice updates the temperature is lowered to a fraction of $0.95$ of the last temperature until a temperature $k_{B} T/J=0.01$ is reached. At each temperature data is collected and $100$ simulations are independently run per set of parameters.

At every temperature the lattice is updated using the Metropolis algorithm followed by a cluster-flipping algorithm. The Metropolis algorithm consists of choosing a spin at random from the lattice and proposing a new direction for this spin. The proposed spin direction is chosen uniformly from the area that is the spherical cap around the original direction where the maximal angle between the original and the proposed spin direction is $\alpha$, and $\alpha$ is dynamically adjusted so that the acceptance probability is approximately 50 \%. The cluster-flipping algorithm grows a cluster of spins and flips all the spins in the opposite direction similar to the Wolff algorithm for Ising spins. In both cases acceptance rates for changing a spin (or cluster of spins) are based on the energy difference between the spin states before and after the move such that the system is sampled according to the Boltzmann distribution.\cite{barkema1}

\begin{center}
\begin{figure}[h!] \centering
\epsfig{file=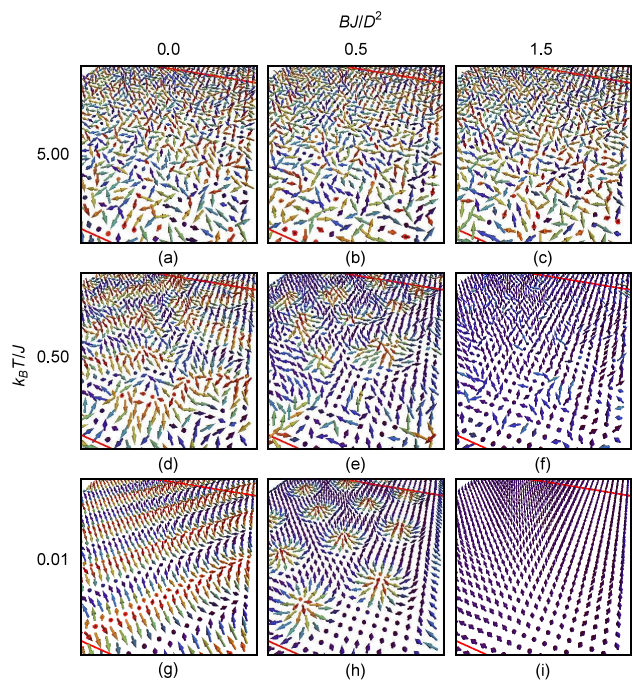,width=1.0\linewidth,clip=}
\caption{Simulations snapshots of spin systems of size $L \times L = 32 \times 32$ for different parameter values and temperatures with the parameter $D/J$ chosen such that the pitch length $p=8$. The periodic boundary is drawn as a  red solid line and spin vectors are represented by  colored arrows where the color scales linearly with the ${\bf \hat{z}}$-component of the vector to highlight spin textures. From top to bottom $k_{B} T/J$ takes on values $5.00$, $0.50$, and $0.01$ respectively and the system undergoes a transition from an unpolarized state to an ordered state in this temperature range. In all cases $K J / D^2$ is zero and from left to right $B J / D^2$ has values $0.0$, $0.5$, and $1.5$, resulting in a spiral, skyrmion and polarized state, respectively, for low enough temperatures.}
    \label{fig:phases}
\end{figure}
\end{center}

\section{Magnetic phases and phase diagram \label{sec:pd}}
Throughout this article we use a nonstandard definition of phases and phase diagram. In the systems we consider there can, strictly speaking, not be any thermodynamic phase transition breaking a continuous symmetry for two reasons: (a) we always work at finite temperature in a two-dimensional system where the Mermin-Wagner theorem forbids the breaking of a continuous symmetry even in the thermodynamic limit; (b) we explicitly consider finite size systems where a real thermodynamic phase transition is also ruled out. Nonetheless, it is possible to identify phases that are distinct in their magnetic configurations, such as phases that may or may not have magnetic skyrmions.

In order to distinguish such phases, a useful quantity is the winding number $w$, which plays the role of a topological charge. It is quantized in a system with periodic boundary conditions and in terms of the unit vector ${\bf{n}}=\langle {\bf{S}} \rangle/|\langle {\bf{S}} \rangle |$ (we choose ${\bf n}$ instead of ${\bf{S}}$ to distinguish between the microscopic spin and the order parameter field) it reads
\begin{align}\label{eq:windingnumber1}
w =
	\frac{1}{4 \pi} \int dx dy {\bf n} \cdot \left(  \partial_{x} {\bf n} \times \partial_{y} {\bf n}  \right)~,
\end{align}
in the continuum limit.  In the definition of ${\bf n}$ and throughout this article angular brackets denote the expectation value in the canonical ensemble. A single (anti-) skyrmion contributes (minus) one to the winding number. This effectively means that the number of skyrmions in an ordered state can be counted. From a set of independent simulations with identical parameters the susceptiblity of the winding number  $\chi_w = \left( \langle w^2 \rangle - \langle w \rangle^2\right) J/k_{B} T$ is calculated. In the latter expression the expectation values are determined by using Eq.~(\ref{eq:windingnumber1}), with ${\bf n}$ replaced by ${\bf S}$ and the integral replaced by a sum, for a given spin configuration in the simulations and then averaging over many such configurations. Whenever the system undergoes a transition between states with different number of skyrmions this susceptibility will be enhanced. 

To determine the phase diagram we take the linear system size $L$ equal to $16$ with one periodic boundary condition (i.e., a wire geometry) at fixed parameter values  $B J / D^2$ and $K J / D^2$. We note here that we simulate a square lattice for convenience throughout, and have not performed finite-size scaling as a function of wire length as we expect our findings to merely change quantitatively for larger system size in the periodic direction. We slowly cooled down the system below the critical temperature and measured the susceptibility of the winding number $\chi_w$. For $k_{B} T/J=0.5$ and zero temperature we find the phase diagram in Fig. \ref{fig:phasediagram}, divided into an easy-axis ($K<0$) and easy-plane ($K>0$) part.  Within the solid line in this figure, the average winding number is larger than one half, so that magnetic skyrmion configurations are expected. The dashed line corresponds to the zero temperature case for the same wire-like confined geometry. The dotted  line corresponds to the infinite system at zero temperature. \cite{leonovphd} Below the skyrmion phase boundary one finds a phase where spiral states rather than skyrmions are stabilized in the infinitely large system. At elevated temperatures ($k_B T/J=0.5$) and in the confined geometry we also find spiral states for fields below the skyrmion phase, albeit that more complicated textures also appear (depending on system size, pitch length, and anistropy;  below we discuss the magnetic configurations in this region in more detail). In the confined geometry the skyrmion phase becomes larger (with respect to the infinite system) at the expense of the spiral phase. We attribute this to the larger ability of skyrmions as compared to spirals to adapt to the repulsive forces away from the edges of the system. \cite{rohart2013} Du {\it et al.} have observed this enhanced stability for Bloch skyrmions in FeGe nanowires. \cite{du2015b}

For fields too large to stabilize skyrmions, the spins are uniformly polarized. In this part of the phase diagram and for fields larger than $2K$ (with $K>0$) the spins are pointing along the field, i.e., along the $z$-direction. For smaller fields and $K>0$ the easy-plane anisotropy tilts the spins away from the field direction. 

The colors in the phase diagram indicate the susceptibility of the winding number. Depending on system size relative to pitch length, the magnetic configuration within the skyrmion phase may undergo transitions between phases with different number of skyrmions. At each such transition the winding number susceptibility is enhanced.

\onecolumngrid
\begin{center}
	\begin{figure}[h!] \centering
		\epsfig{file=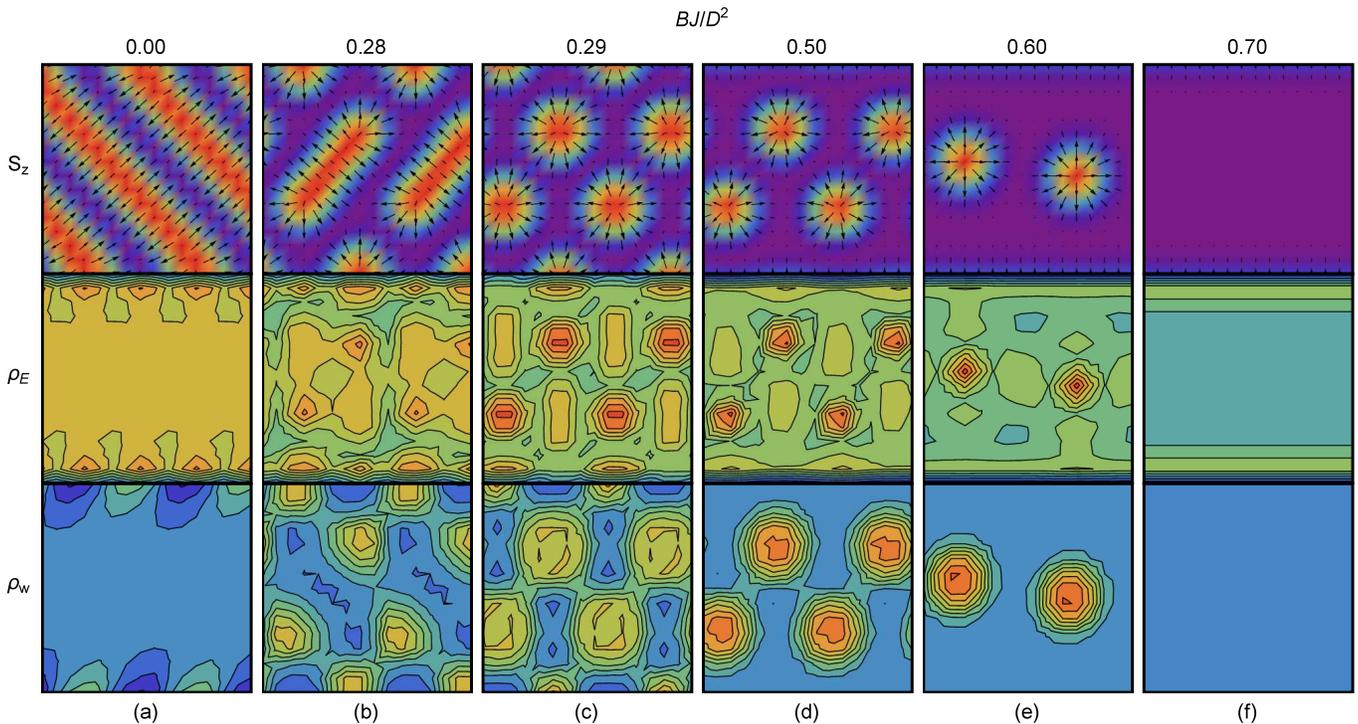,width=1.0\linewidth,clip=}
		\caption{Magnetization (top row), energy density $\rho_E$ (middle row), and topological charge density $\rho_w$ (bottom row) for different values of the field for a square system with linear size $L=16$ with periodic boundaries in the horizontal direction. The pitch lengh is $p=7$ and the temperature is zero. The color coding indicates out of plane magnetization (top row) and energy and topological charge densities (middle and bottom row). }
		\label{fig:cascade}
	\end{figure}
\end{center}
\twocolumngrid

Finally, we note that the phase diagram at zero temperature was studied before  for infinite system size  with different methods. \cite{randeria1} Contrary to this latter work we do not find a tri-critical point where polarized, spiral and skyrmion phases meet in the easy-plane part of the phase diagram from our simulations. 
Representative spin configurations of the polarized state, the spiral state, and the skyrmion state at small temperature can be found in Fig.~\ref{fig:phases} (g), (h), and (i), respectively. We also note that for Bloch skyrmions the role of anisotropy on the ground-state phase diagram was discussed in Ref.~\onlinecite{wilson2014}.

As we have already mentioned in the discussion of the phase diagram, the system goes through different magnetic configurations upon lowering the field through the skyrmion phase to zero. The precise configuration depends on the ratio of system size to pitch length. In Fig.~\ref{fig:cascade} we show the situation for a wire geometry with linear size $L=16$ and one periodic boundary at zero temperature and for zero anisotropy. Starting from large fields (and hence from the polarized phase), at a certain field strength (see Fig.~\ref{fig:phasediagram}) the magnetic configuration changes from being uniformly polarized, into a configuration with skyrmions, upon lowering the field. As the distance between skyrmions depends on the field (with larger field leading to larger distance) the number of skyrmions increases discontinuously as the field is lowered. Roughly speaking, the number of skyrmions jumps once the field (and thus preferred skyrmion distance) is low enough to accommodate more skyrmions in the confined geometry. Note that this clearly depends on system size. For lower fields, the skyrmion configuration changes into a spiral state via a state where the skyrmions are elongated. Such extended skyrmions are reminiscent of ``fingers'' in liquid crystals that appear in many varieties. \cite{bogdanov1, ackerman2014, leonov2014} Also note that for low enough fields  half skyrmions appear  at the edge of the system. For very low fields, the spiral state is stabilized and in the middle of the system the elongated skyrmions and spirals orient themselves $90$ degrees with respect to the lattice to maximize the period of the spiral (and thus minimize exchange energy). Similar anisotropies may very well be present in some  materials and in our simulations result from the underlying lattice. For simulations that would need to give quantitative predictions for continuum systems, one could add additional terms that make the exchange interactions more isotropic. \cite{buhrandt1} Here, we do not pursue this route as we are interested in the qualitative features of the phases and cascade of phase transitions. At the edge of the wire the influence of exchange is less important with respect to DM interactions and the spirals are parallel to the edge. Because of the strong easy-plane anisotropy of MnSi thin films, Wilson {\it et al.} considered helicoids with wave vector perpendicular to the edges, and, moreover, only considered edge states without internal structure. \cite{wilson2013} In the geometry we consider, the edge states may also consist of half skyrmions, and, as a result, the spiral states that finger out of the skyrmions and half skyrmions upon lowering the field will have their wave vector parallel to the edges of the wire. Again, we note that several of the features we dicuss, such as half and elongated skyrmions, and spiral orientation at the edges, were very recently experimentally observed by Du {\it et al.} for Bloch skyrmions. \cite{du2015b}

In Fig.~\ref{fig:cascade} we show, in addition to the magnetization in the top row, the energy density $\rho_E$ [the expectation value of the summand in Eq.~(\ref{eq:hamiltonian})] in the middle row of figures, and the topological charge (winding number) density  $\rho_w$ [integrand in Eq.~(\ref{eq:windingnumber1})]  in the bottom row, where the brighter colors indicate higher values. At the skyrmion (and half skyrmion) positions the energy is largest as the spin at the skyrmion core points opposite to the external field. We also note that the figure of the magnetization clearly shows edge states where the magnetization tilts away from the field direction at the boundaries of the system. \cite{rohart2013} The middle row shows that the energy density is minimized at the edges. Note that for low enough field the half skyrmions at the edge are formed from these edge states, as was also discussed in Ref.~\onlinecite{meynell2014} for Bloch skyrmions. We note that in structures where half skyrmions appear the edges show alternating positive and negative contributions to the winding number, while in the skyrmion phase the contributions to the winding number are positive and due to the skyrmions only.

\section{Degeneracies and fluctuations \label{sec:degen}}
\begin{center}
\begin{figure}[h!] \centering
\epsfig{file=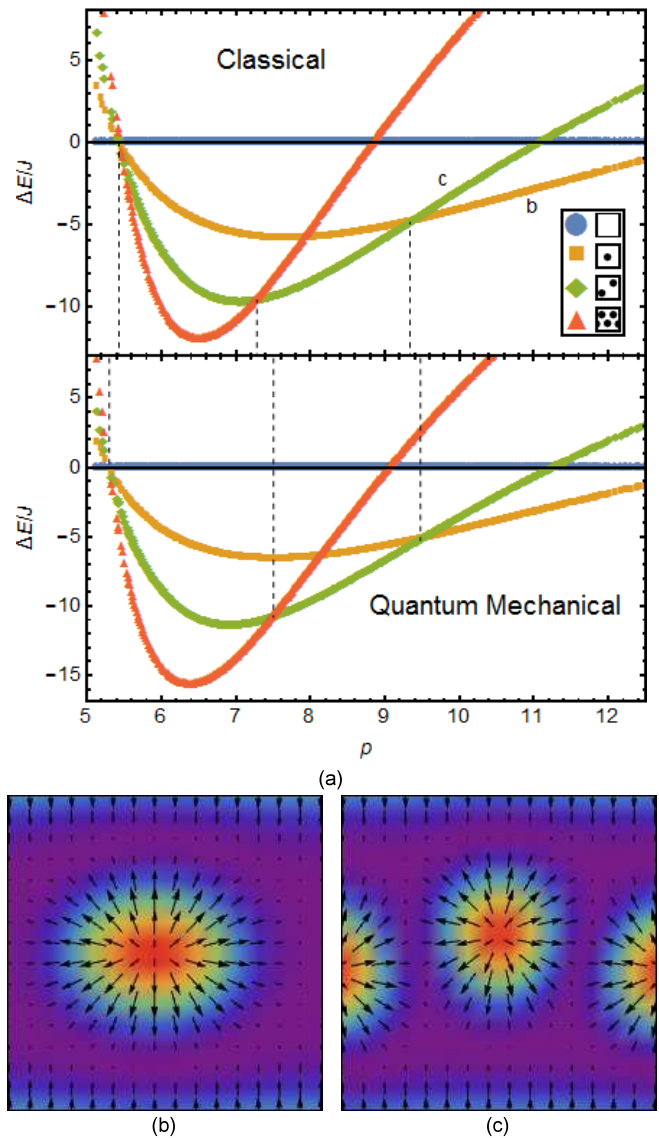,width=1.0\linewidth,clip=}
\caption{(a): Difference in energy (upper panel: classical energy, lower panel: energy including quantum correction) between configurations with skyrmions and the state without skyrmions at zero temperature versus pitch length, for various numbers of skyrmions and system size $L\times L=16\times16$ with one periodic and one open boundary and parameters $K J/D^2=0.0$ and $B J /D^2=0.5$. The blue circular, yellow rectangular, green diamond-shaped, and red triangular data points correspond to systems with 0, 1, 2, and 4 skyrmions respectively. (b) and (c): Magnetic configuration of two classically-degenerate ground states with energy $E=-596.17 J$ of a system at pitch length $p=9.336$ deep in the skyrmion regime with one periodic (horizontal direction) and one open (vertical direction) boundary.}
    \label{fig:energyvspitch}
\end{figure}
\end{center}

At the transitions between different magnetic configurations as a function of field, two magnetic configurations are degenerate. Such degeneracies occur generically in a confining geometry as a function of field, ratio of DM to exchange interaction, or system size, because the geometry prevents the skyrmions from reaching their preferred (in an infinitely large system) position. This leads, at some particular set of values of the parameters, e.g. to a degeneracy between a state with fewer but larger skyrmions and a state with more but smaller skyrmions.  An example of two degenerate spin configurations is given in Fig.~\ref{fig:energyvspitch}. In this figure, the skyrmions are elliptical as the edge states effectively push the skyrmions away from the boundaries \cite{rohart2013} thereby deforming the skyrmions. Skyrmion deformation due to anisotropy was discussed in Ref.~\onlinecite{wilson2012}. We now turn to the question of how fluctuations lift zero-temperature degeneracies between two magnetic configurations, in particular for degenerate configurations containing a different number of skyrmions. 

To investigate this in detail we look again at a wire geometry of size $L\times L=16\times16$ with $K J/D^2=0.0$ and $B J /D^2=0.5$. We vary the pitch length $p$ and determine the classical energy of different skyrmion configurations. These energies are found by initiating the simulations with a certain number of skyrmions and using simulated annealing to force the configuration to an energetic local minimum at zero temperature. The upper panel of Fig.~\ref{fig:energyvspitch}~(a) shows the classical ground-state energies for various configurations as a function of the pitch length relative to the energy of a system without any skyrmions. There is a range of pitch lengths $5.5 \lesssim p \lesssim 13.5$ for which the classical ground state is a configuration containing skyrmions.

We note that a similar result can be obtained by varying the system size rather than the pitch length, since the ratio between these two length scales determines the preferred amount of skyrmions in the system. This means that for a given material with certain values for $J$, $D$, and $K$ and some applied magnetic field such that skyrmions are expected, wires can be made with a certain thickness so that their classical magnetic ground state is degenerate. 

To investigate the effect of quantum and thermal spin-wave (magnon) fluctuations at zero and nonzero temperature, respectively, we use the method outlined in Refs.~[\onlinecite{roldan2014}] and [\onlinecite{roldan2015}]. We quantize the spins and use a Holstein-Primakoff transformation to bosonic operators $\hat a_{\bf r}$ and $\hat a_{\bf r}^\dagger$. It is given by
\begin{eqnarray}
\label{eq:hptrafo} 
 \hat {\bf S}_{\bf r} \cdot \bm{\Omega}_{\bf r} &=& S - \hat  n_{\bf r}~; \nonumber \\
 \hat  S^-_{\bf r} &=& \hat a_{\rm r} \sqrt{2S - \hat n_{\bf r}}~,
\end{eqnarray}
where $ \hat  S^-_{\bf r}$ is the usual spin-lowering operator and $\hat n_{\bf r} = \hat a^\dagger_{\bf r} \hat a_{\bf r}$. Moreover,   $\bm{\Omega}_{\bf r}$ denotes the classical spin configuration that is found from the simulations at zero temperature and $S$ is the spin quantum number (which we take equal to one as the simulations are done for normalized spins). We insert the above transformation in the hamiltonian and keep terms up to quadratic order in the creation and annihilations operators which amounts to a linear approximation in which interactions between spin waves are neglected, which is sufficient for low temperatures. The hamiltonian acquires terms $\sim \hat a \hat a$ and $\sim \hat a^\dagger \hat a^\dagger$ that are removed by a Bogoliubov transformation to new bosonic operators $\hat \gamma_i^\dagger$ and $\hat \gamma_i$ that respectively create and annihilate a spin wave with energy $\epsilon_i$. Here, $i$ is an index that labels the spin-wave modes. After the Bogoliubov transformation, the hamiltonian is in the above-mentioned harmonic approximation given by \cite{roldan2014}
\begin{equation}
\label{eq:quantumham}
  H = E_{\rm cl} + E_0 + \sum_i  \epsilon_i \left(\hat \gamma_i^\dagger \hat \gamma_i + \frac{1}{2} \right)~.
\end{equation}
The first term is the classical ground-state energy as found in the simulations whereas the second term is a quantum contribution that arises in the Bogoliubov approach outlined aboved. This latter contribution is absent at the classical level. 

Using the above hamiltonian, the ground-state energy including quantum corrections is found to be
\begin{equation}
\label{eq:quantumenergy}
    E  = E_{\rm cl} + E_0 + \sum_i   \frac{\epsilon_i }{2}~.
\end{equation}
At a point where two magnetic configurations  in our simulations are found to be degenerate, the first term $E_{\rm cl}$ is equal for them. The spin-wave spectrum (and hence the quantum correction represented by the last two terms in the above expression for the energy) is, however, generically different for two classically degenerate configurations, so that quantum fluctuations may indeed remove classical degeneracies. In the lower panel of Fig.~\ref{fig:energyvspitch}~(a) we show the energy including quantum corrections for magnetic configurations containing up to four skyrmions as a function of pitch length and for the same parameters as the upper panel of this figure. This result shows that the energies are shifted by the quantum corrections. Moreover, they are shifted in such a way that the region where the configuration with four skyrmions is the true lowest-energy state is enlarged with respect to the classical result. This conclusion is in line with the finding of Rold\'an-Molina {\it et al.} [\onlinecite{roldan2015}] that quantum fluctuations stabilize skyrmion textures over the collinear ferromagnetic phase. Loosely speaking this comes about because larger gradients in the spin direction lead to more quantum fluctuations.

Having discussed the effect of quantum, i.e., zero-temperature, fluctuations, we now turn to thermal fluctuations. To investigate which configuration is entropically preferred, we i) compute the entropy due to the spin waves around two degenerate skyrmion configurations (an approach appropriate at low temperatures), and ii) directly measure the relative probability of occurrence of configurations at nonzero temperatures within our simulations (and approach valid at intermediate and high temperatures). In the first approach the spin waves (or magnons) are considered as non-interacting bosonic particles thus taking into account their quantum statistics. We refer to this approach as the spin-wave analysis. 

\begin{center}
\begin{figure}[h!] \centering
\epsfig{file=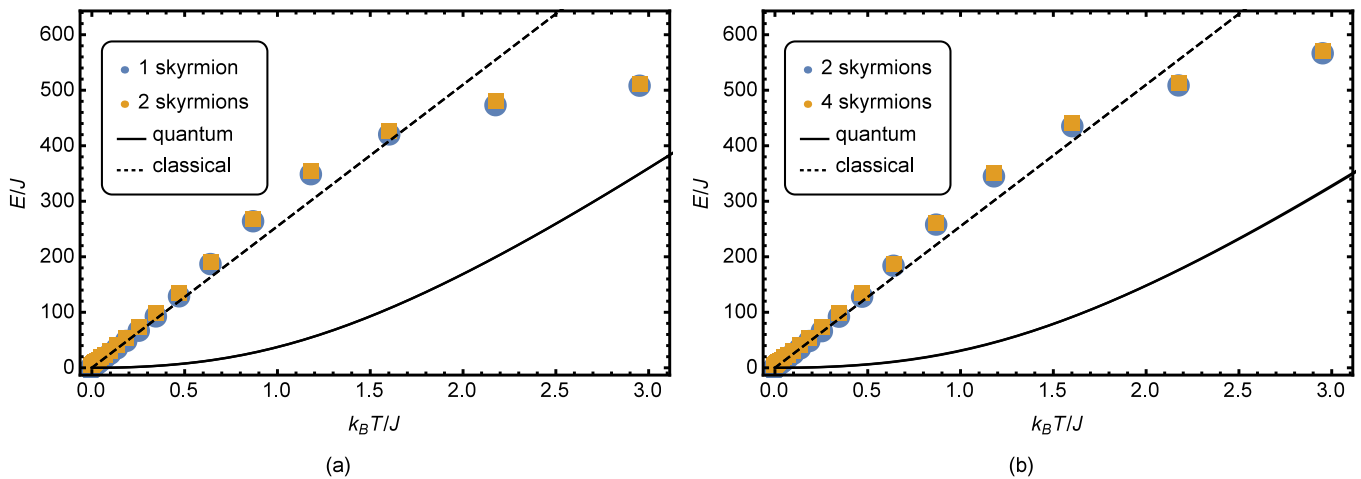,width=1.0\linewidth,clip=}
\caption{The energyof various spin configurations, with pitch length $p=9.336$ in (a) and $p=7.284$ in (b), as a function of temperature $k_{B} T / J$ and where $E_{\rm cl}$ is put equal to zero. In both cases the parameters are $K J/D^2=0.0$ and $B J /D^2=0.5$. The data points represent data from Monte Carlo simulations that are in agreement with results from the equipartition theorem at low temperatures (shown by the black dashed line). The solid black lines represent the energies resulting from the spin-wave analysis.}
    \label{fig:energyvst}
\end{figure}
\end{center}

We look at a system with $p=9.336$ and $p=7.284$. At these values of the pitch length, the classically degenerate ground states contain either one or two skyrmions for $p=9.336$ as in Fig.~\ref{fig:energyvspitch} or two or four skyrmions for $p=7.284$.  At nonzero temperature, the total energy of the system is  in the harmonic approximation given by 
\begin{align}\label{eq:energyBE}
E (T)=E_{\rm cl} + \sum_i \epsilon_i n_i~,
\end{align}
with the Bose-Einstein distribution function $n_i=[\exp(\beta \epsilon_i)-1]^{-1}$. In the above equation and following discussion we neglect the quantum corrections to the energy discussed previously. The spin-wave entropy is given by
\begin{align}\label{eq:entropyBE}
S (T)= - k_{B} \sum_i \left[ n_i \ln n_i -(1+n_i)\ln(1+n_i) \right]~.
\end{align}
Since the simulations are classical we expect them to correspond to the Rayleigh-Jeans limit, $n_i \to 1/\beta \epsilon_i$, of the above formulas, which leads to equipartition of energy such that in this limit the total energy is equal to $N k_B T$, with $N$ the number of spins. In Fig.~\ref{fig:energyvst} we display the average energies of the degenerate ground states as a function of temperature. As expected, at low temperatures the simulation agrees with this classical equipartition result, whereas the quantum-mechanical result is suppressed with respect to equipartition. This is because apart from a few (nearly) zero modes, most spin-wave excitations have energies $\sim B$, such that at temperatures $k_B T \ll B$ our simulations overestimate their contribution to the energy. The deviation of the simulations from equipartition at high temperatures is because in this limit the harmonic approximation starts to break down. 

\onecolumngrid
\twocolumngrid

\onecolumngrid
\begin{center}
\begin{figure}[h!] \centering
\epsfig{file=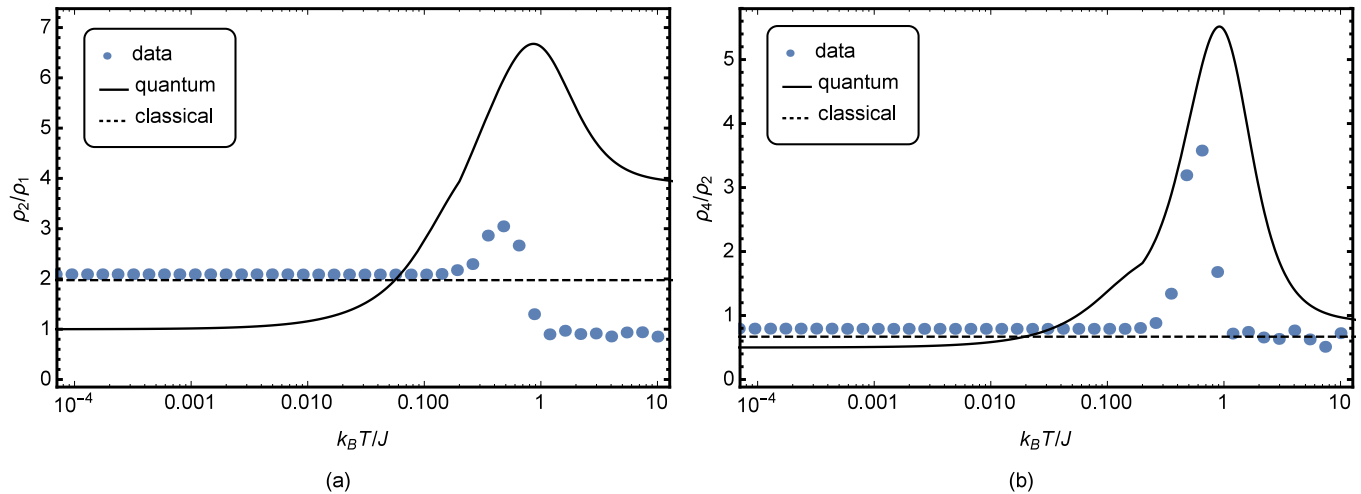,width=1.0\linewidth,clip=}
\caption{Ratio of probabilities of having a certain amount of skyrmions for two different situations with parameters $K J/D^2=0.0$ and $B J /D^2=0.5$ and pitch length $p=9.336$ for  (a) and $p=7.284$ for (b) as a function of temperature. The dots correspond to results from the Monte Carlo simulations, the solid-line from the spin-wave analysis not including translation entropy, and the dashed line to the zero-temperature classical limit.}
    \label{fig:ratiovst}
\end{figure}
\end{center}
\twocolumngrid

At nonzero temperatures, the probability of the system being in one out of two degenerate ground states depends on their difference in entropy. Besides the entropy due to magnons there is also  translational entropy that depends on the number of skyrmions, and that needs to be included in the overall entropy of a configuration. The simulations automatically include this extra entropy, but it is not accounted for in the expression in Eq.~(\ref{eq:entropyBE}) and needs to be included on top of this expression. For example, the skyrmion configurations in Fig.~\ref{fig:energyvspitch} can be translated in the periodic direction by a lattice constant. For a single skyrmion there are $16$ (for system size 16 in the periodic direction) translations, whereas for two skyrmions the configuration can be mapped onto itself by a translation over $8$ lattice site in combination with a reflection. Hence, the configurations with one and two skyrmions have the same translational entropy. Similar counting leads to the conclusion that there are twice as much translations possible for the situation of two, as compared to four skyrmions so that the translational entropy of the configuration with two skyrmions is $k_B \ln 2 $ higher than that of four skyrmions. 

Fig.~\ref{fig:ratiovst} displays the ratio of probabilities for occurence of skyrmion configurations with different number of skyrmions: two and one skyrmions for $p=9.336$, and four and two skyrmions for $p=7.284$, as a function of temperature. In this figure, both simulation results (dots) and results from the spin-wave analysis without  the translational entropy are shown (black solid line). The probability that results from the classical limit of the entropy [Eq.~(\ref{eq:entropyBE}) with the replacement $n_i \to 1/\beta \epsilon_i$, dashed black line] leads to a probability ratio that agrees with the low-temperature limit of our simulations, where the ratio was measured by cooling a system down $10^4$ times while measuring the number of skyrmions in the system. According to the spin-wave analysis, the probability ratio is in the low-temperature only determined by translation entropy. Both the quantum-mechanical result for the entropy as well as the simulations show a peak around the ferromagnetic phase transition $k_{B} T / J\sim 1$. At moderate temperatures ($0.1 J -0.5 J$) our results show that the configuration with the most skyrmions is entropically favored.  

\begin{center}
\begin{figure}[h!] \centering
\epsfig{file=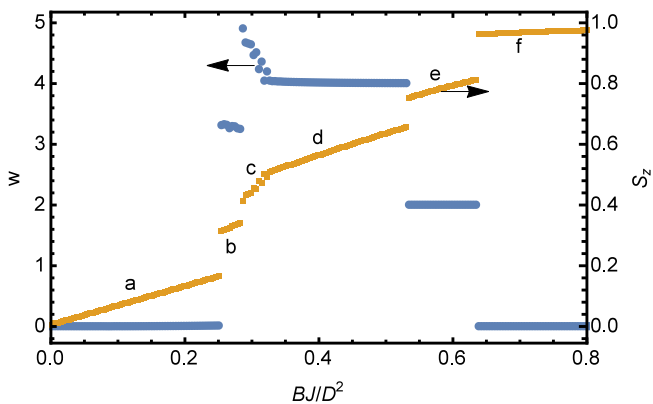,width=1.0\linewidth,clip=}
\caption{Winding number (left axis) and average magnetization in the $z$-direction (right axis) as a function of magnetic field for the same parameters as Fig.~\ref{fig:cascade}. Labels a-f refer to states in Fig.~\ref{fig:cascade}. All parameters are taken the same as for the results Fig.~\ref{fig:cascade}.}
    \label{fig:wandszvsb}
\end{figure}
\end{center}

\section{Conclusions, discussion, and outlook \label{sec:conc}}
In this article we have shown that in finite systems and at elevated temperatures skyrmions are present in a large part of the phase diagram. We have also discussed how the magnetic field tunes the system through a cascade of transitions between different magnetic configurations, and how zero-temperature degeneracies between such magnetic configurations are lifted by fluctuations. Throughout this article we have focused on PMA materials where the DM interactions are believed to arise due to interfaces between very thin layers of magnetic materials (such as Co)  and materials with strong spin-orbit couping (such as Pt).  \cite{emori1,ryu1,ryu2,franken1}  Such DM interactions give rise to N\'e\`el skyrmions. Because the magnetic layers in these system are very thin (only a few atoms), our two-dimensional treatment is appropriate. In particular, the two-dimensional nature and form of the DM interactions prevent the formation of conical phases in the phase diagram. This situation is different from the situation of thin films of MnSi in which Bloch skyrmions are stabilized and conical phases are present.\cite{wilson2012, wilson2013}

In this article we have focused on the confined geometry of a wire. The cascade of transitions as a function of field is generic and appears for any confined geometry because the field influences the preferred skyrmion distance. Which particular magnetic structures occur in the cascade depends on the confined geometry, however, and on the form of the DM interactions and anisotropy. In particular, we expect that the half-skyrmions that we have found will only appear in a wire geometry. 

Jumps in the magnetoresistance in MnSi nanowires have been observed recently, and were attributed to changes in the number of skyrmions in the magnetic configuration. \cite{du2015} Measurements of  changes in the topological Hall effect due to the appearance of single skyrmions in a confined geometry were performed on FeGe. \cite{kanazawa2015} Motivated by these experiments we now investigate whether the transitions between various configurations that appear in our system as a function of field (see  Fig.~\ref{fig:cascade}) can be detected electrically.  To this end, we compute the total winding number and the total magnetization in the $z$-direction as a function of field for the same parameters as Fig.~\ref{fig:cascade} (see Fig.~\ref{fig:wandszvsb}). The total winding number determines (up to prefactors) the topological Hall signal, provided spin-orbit coupling is small. \cite{lee1, neubauer1} The result in Fig.~\ref{fig:wandszvsb} clearly shows jumps in the winding number as the magnetic configuration undergoes a structural transition. For PMA materials, however, the topological Hall signal is expected to be very small. \cite{knoester2014} An alternative for electrical detection of the magnetization configuration is then the anomalous Hall signal that is proportional to the total magnetization in the $z$-direction for our geometry. Fig.~\ref{fig:wandszvsb} shows that this quantity also jumps as the magnetization configuration undergoes a transition as a function of field. Based on this analysis we conclude that the transitions between various magnetic configurations that have we found may be detected electrically. 

Using experimental parameters for PMA materials from Ref.~\onlinecite{emori1} we estimate that  $K J / D^2   \sim -1$ and $B J / D^2   \sim 10^{-4} - 10^{-1}$ for these experiments.  The route to observe skyrmions in these systems would therefore be to increase the field and lower the anisotropy (or preferrably make it easy-plane). Very recently, Moreau-Luchaire {\it et al.} have reported the observation of skyrmions at room temperature in multilayers of Co and Pt with PMA. \cite{moreau2015} The skyrmions observed in these measurements are rather large and stabilized as a result of both dipole-dipole and DM interactions, and are therefore in a somewhat different regime from the skyrmions that we have studied in this article. 

In future work, we will investigate how current-induced torques manipulate the skyrmionic magnetic structures we have found. Finally, motivated by the recent experimental results of Du {\it et al.} \cite{du2015b} we also intend to consider Bloch skyrmions. 

\section*{Acknowledgement \label{sec:ack}}
It is a pleasure to thank Gerrit Bauer, Reinoud Lavrijsen and Henk Swagten for useful comments. LF and SB acknowledge support from the Deutsche Forschungsgemeinschaft under FR 2627/3-1. This work is supported by the Stichting voor Fundamenteel Onderzoek der Materie (FOM) and is part of the D-ITP consortium, a program of the Netherlands Organisation for Scientific Research (NWO) that is funded by the Dutch Ministry of Education, Culture and Science (OCW).

\end{document}